\documentclass[twocolumn,prd,preprintnumbers,amsmath,amssymb]{revtex4}

\usepackage{graphicx}
\usepackage{dcolumn}
\usepackage{bm}
\usepackage{epsfig}
\usepackage{amssymb}
\usepackage{amsmath}

 \def\tskip{\setlength{\tskip}{5pt}}
\def\colwidth{\setlength{\colwidth}{3.5in}}

\newcommand{\lsim}{\mathrel{\hbox{\rlap{\lower.55ex\hbox{$\sim$}} \kern-.3em \raise.4ex \hbox{$<$}}}}
\newcommand{\gsim}{\mathrel{\hbox{\rlap{\lower.55ex\hbox{$\sim$}} \kern-.3em \raise.4ex \hbox{$>$}}}}
\newcommand{\be}{\begin{equation}}
\newcommand{\ee}{\end{equation}}
\newcommand{\ba}{\begin{eqnarray}}
\newcommand{\ea}{\end{eqnarray}}

\begin{document}

\title{Attractor Solution in Coupled Yang-Mills Field Dark Energy Models}

\author{Wen Zhao\footnote {wzhao7@mail.ustc.edu.cn}}
\affiliation{Department of Physics, Zhejiang University of Technology, Hangzhou, 310014, P. R. China\\School of Physics and Astronomy, Cardiff University,
Cardiff CF24 3AA, United Kingdom  }

\date{\today}

\begin{abstract}
We investigate the attractor solution in the coupled Yang-Mills field
dark energy models with the general interaction term, and obtain the constraint equations for the interaction if the attractor solution exists. The research also shows that, if the attractor solution exists, the equation-of-state
of the dark energy must evolve from $w_y>0$ to $w_y\le-1$, which is slightly suggested by the observation.
At the same time, the total equation-of-state in the attractor
solution is $w_{tot}=-1$, the universe is a de Sitter expansion,
and the cosmic big rip is naturally avoided. These features are
all independent of the interacting forms.\end{abstract}


\maketitle

\section{Introduction \label{introduction}}

Dark energy problem has been one of the most active fields in the
model cosmology, since the discovery of accelerated expansion of
universe\cite{sn,map,sdss}. In the observational cosmology, the
equation-of-state (EOS) of the dark energy $w_{de}\equiv
p_{de}/\rho_{de}$ plays a central role, where $p_{de}$ and
$\rho_{de}$ are its pressure and energy density, respectively. To
accelerate the expansion, the EOS of dark energy must satisfy
$w_{de}<-1/3$. The simplest candidate of dark energy is a tiny
positive time-independent cosmological constant $\Lambda$, whose
EOS is $-1$. However, it is difficult to understand why the
cosmological constant is about $120$ orders of magnitude smaller
that its natural expectation, i.e. the Planck energy scale
density. This is the so-called fine-tuning problem.
Another puzzle of the dark energy is the first cosmological
coincidence problem\cite{coin}, namely, \emph{why does our
universe begin the accelerated expansion recently? why are we
living in an epoch in which the dark energy density and the dust
matter energy density are comparable?} This problem becomes very
serious especially for the cosmological constant as the dark
energy candidate. The cosmological constant remains unchanged
while the energy densities of dust matter and radiation decrease
rapidly with the expansion of our universe. Thus, it is necessary
to make some fine-tuning. In order to give a reasonable
interpretation to the first cosmological coincidence problem, many
dynamical dark energy models have been proposed as alternatives to
the cosmological constant, such as quintessence\cite{quint},
phantom\cite{phantom}, k-essence\cite{k}, quintom\cite{quintom}
etc.

Recently, by fitting the SNe Ia data, marginal evidence for
$w_{de}(z)<-1$ at redshift $z<0.2$ has been found. In addition,
many best fits of the present values of $w_{de}$ are less than
$-1$ in various data fittings with different parameterizations.
The present observational data seem to slightly favor an evolving
dark energy with $w_{de}$ crossing $-1$ from above to below in the
near past\cite{trans}. In has been found that the EOS of dark
energy $w_{de}$ cannot cross the so-called phantom divided
$w_{de}=-1$ for quintessence, phantom or k-essence
alone\cite{vikman}. A number of works have discussed the quintom
models\cite{quintom}, which is an combination of a quintessence
and a phantom. Although many of these models provide the
possibility that $w_{de}$ can cross $-1$, they do not answer
another question, namely, \emph{why crossing phantom divided
occurs recently?} Since in many existing models whose EOS can
cross the phantom divide, $w_{de}$ undulated around $-1$ randomly,
why are we living in an epoch $w_{de}<-1$? It is regarded as the
second cosmological coincidence problem\cite{sc}.

As well known, the most frequently used approach to alleviate
the first cosmological coincidence problem is the tracker field dark energy scenario\cite{scaling}. The dark energy can track the evolution of the background matter in the early stage, and only recently, the dark energy has negative pressure, and becomes dominant . Thus the current condition of the dark energy is nearly independent of the initial condition.  If the possible interaction between the dark energy and background matter\cite{couple} is considered, the whole system (including the background matter and dark energy) may be eventually
attracted into the scaling
attractor, a balance achieved, thanks to the interaction.
In the scaling attractor, the effective densities of dark energy
and background matter decrease in the same manner with the
expansion of our universe, and the ratio of dark energy and
background matter becomes a constant. So, it is not strange that
we are living in an epoch when the densities of dark energy and
matter are comparable. In this sense, the first cosmological
coincidence problem is alleviated. On the other hand, if the
scaling attractor also has the property that its EOS of dark
energy is smaller than $-1$, the second cosmological coincidence
problem, if existing, is also alleviated at the same time\cite{sc}. However, this is
impossible in the interacting quintessence or phantom scenario.

Recently, a number of authors have discussed another class of models, which are based on the conjecture that a vector
field can be the origin of the dark energy\cite{vector,vector2}, and have
different features to those of scalar field. In the
Refs.\cite{z,Zhang,zhao1,zhao2}, it is suggested that the
Yang-Mills (YM) field can be a kind of candidate for such a vector
field. Compared with the scalar field, the YM field is the
indispensable cornerstone to particle physics and the gauge bosons
have been observed. There is no room for adjusting the form of
effective YM lagrangian as it is predicted by quantum corrections
according to field theory. In the previous
works\cite{zhao1,zhao2,xia}, we have investigated the 1-loop YM
field case and found attractive features: The YM field dark energy models can naturally realize the EOS of $w_y>-1$
and $w_y<-1$, and the current state of the YM dark energy is independent of the choice of the initial condition. The cosmic big rip is also avoided in the models.

In the recent works \cite{2-loop,3-loop}, the 2-loop and 3-loop YM field dark energy are also considered. Although these cases are much more complicated than the 1-loop case, they have not brought new feature for the evolution of the universe. So in this work, we shall only focus on the YM field with 1-loop case.

In this work, the cosmological evolution of the YM dark energy
interacting with background prefect fluid is investigated. In
fact, gauge fields play a very important role in, and are the
indispensable cornerstone to, particle physics. All known
fundamental interactions between particles are mediated through
gauge bosons. Generally speaking, as a gauge field, the YM field
under consideration may have interactions with other species of
particles in the universe. However, unlike those well known
interaction in QED, QCD, and the electron-weak unification, here
at the moment we do not yet have a model for the details of
microscopic interactions between the YM field and other particles.
In this work, instead of considering some specific assumed interactions between YM field and matter and radiation, which has adopted in \cite{xia,2-loop,3-loop},
we shall consider YM dark energy model with a general interacting term, and investigate the general feature of the attractor solution.

This paper is organized as follows. In Sec.2, we give out the
equations of the dynamical system of the interacting YM field dark
energy models, and discuss the general features of the interacting
models. In Sec.3, we consider three special cases of the
interaction terms and the holographic YM dark energy models, and
investigate the constraints of these interaction terms. Finally,
brief conclusion and discussion are given in Sec.4.


\section{Dynamic system of interacting Yang-Mills dark energy\label{section2}}

The effective YM  field cosmic model has been discussed in
Refs.\cite{z,Zhang,zhao1,zhao2}. The effective lagrangian  up to
1-loop order is \cite{pagels, adler}
 \be
 \mathcal{L}_{eff}=\frac{b}{2}F\ln\left|\frac{F}{e\kappa^2}\right|, \label{L}
 \ee
where  $b=11N/24\pi^2$ for the generic gauge group $SU(N)$ is the
Callan-Symanzik coefficient \cite{Pol}.
$F=-(1/2)F^a_{\mu\nu}F^{a\mu\nu}$ plays the role of the order
parameter of the YM field. $\kappa$ is the renormalization scale
with the dimension of squared mass, the only model parameter. The
attractive features of this effective YM lagrangian include the
gauge invariance, the Lorentz invariance, the correct trace
anomaly, and the asymptotic freedom\cite{pagels}. With the
logarithmic dependence on the field strength, $\L_{eff}$ has a
form similar to he Coleman-Weinberg scalar effective
potential\cite{coleman}, and the Parker-Raval effective gravity
lagrangian\cite{parker}. The effective YM field was firstly put
into the expanding Robertson-Walker (R-W) spacetime to study
inflationary expansion\cite{z} and the dark energy\cite{Zhang}. We
work in a spatially flat R-W spacetime with a metric
 \be
 ds^2=a^2(\tau)(d\tau^2-\delta_{ij}dx^idx^j),\label{me}
 \ee
where $\tau=\int(a_0/a)dt$ is the conformal time. For simplicity
we study the $SU(2)$ group and consider the electric case with
$B^2 \equiv0 $. The energy density and pressure of the YM field
are given by \be
 \rho_y=\frac{E^2}{2}\left(\epsilon+b\right),
 ~~~~p_y=\frac{E^2}{2}\left(\frac{\epsilon}{3}-b\right),
 \ee
where the dielectric constant is given by
 \be
 \epsilon=b\ln\left|\frac{F}{\kappa^2}\right|,\label{epsilon}
 \ee
and the EOS is
 \be
 w_y=\frac{p_y}{\rho_y}= \frac{y-3}{3y+3},\label{13}
 \ee
where $y\equiv\epsilon/b=\ln|\frac{F}{\kappa^2}|$. At the critical
point with the order parameter $F=\kappa^2$, one has $y=0$ and
$w_y=-1$, the universe is in  exact de Sitter expansion \cite{z}.
Around this critical point, $F< \kappa^2$ gives $y<0$ and
$w_y<-1$, and $F> \kappa^2$ gives $y>0$ and $w_y>-1$. So in the YM
field model, EOS of $w_y
>-1$ and $w_y<-1$ all can be naturally realized. When
$y\gg1$, the YM field has a state of $w_y=1/3$, becoming a
radiation component. The effective YM equations are
 \be
 \partial_{\mu}(a^4\epsilon~
 F^{a\mu\nu})+f^{abc}A_{\mu}^{b}(a^4\epsilon~F^{c\mu\nu})=0,
 \label{F1}
 \ee
the $\nu=0$ component of which is an identity, and the $\nu=1,2,3
$ spatial components of which reduce to
 \be\label{ymequation}
 \partial_{\tau}(a^2\epsilon E)=0.
 \ee

In this work we will generalize the original YM dark energy model
to include the interaction between the YM dark energy and dust
matter. We assume the YM dark energy and background matter
interact through an interaction term $Q$, according to
 \be
 \dot{\rho}_y+3H(\rho_y+p_y)=-Q,\label{doty}
 \ee
 \be
 \dot{\rho}_m+3H\rho_m=Q,\label{dotm}
 \ee
which preserves the total energy conservation equation
$\dot{\rho}_{tot}+3H(\rho_{tot}+p_{tot})=0$. It is worth noting
that the equation of motion (\ref{ymequation}) should be changed
when $Q\neq0$. We introduce the following dimensionless variables
 \be
 x\equiv\frac{2\rho_m}{b\kappa^2},~~f\equiv\frac{2Q}{b\kappa^2H},
 \ee
where $f$ is the function of $x$ and $y$. By the help of the
definition of $y$, the evolution equations (\ref{doty}) and
(\ref{dotm}) can be rewritten as a dynamical system, i.e.
 \ba
 y'&=&-\frac{4y}{2+y}-\frac{f(x,y)}{(2+y)e^y},\label{y'}\\
 x'&=&-3x+f(x,y),\label{x'}
 \ea
here, a prime denotes derivative with respect to the so-called
e-folding time $N\equiv\ln a$. The fractional energy densities of
dark energy and background matter are given by
 \be
 \Omega_y=\frac{(1+y)e^y}{(1+y)e^y+x},~~{\rm
 and}~~\Omega_{m}=\frac{x}{(1+y)e^y+x}.\label{Omega}
 \ee

We can obtain the critical point $(y_c,x_c)$ of the autonomous
system by imposing the conditions $y_c'=x_c'=0$. From the
equations (\ref{y'}) and (\ref{x'}), we obtain that the critical
state satisfies the following simple relations
 \be
 3x_c=f(x_c,y_c),\label{[1]}
 \ee
 \be
 3x_c=-4y_ce^{y_c},\label{[2]}
 \ee
so we can get the critical state $(y_c,x_c)$ by solving these two
equations. In order to study the stability of the critical point,
we substitude linear perturbations $y\rightarrow y_c+\delta y$ and
$x\rightarrow x_c+\delta x$ about the critical point into
dynamical system equations (\ref{y'}) and (\ref{x'}) and linearize
them, and obtain two independent evolutive equations, i.e.
\[
\left(
 \begin{array}{c}
 \delta y'\\
 \delta x'
  \end{array}
 \right)
 \equiv
M \left(
 \begin{array}{c}
 \delta y\\
 \delta x
  \end{array}
 \right)  =
\left( \begin{array}{cc}
G_{\rm y}+R_{\rm y} & R_{\rm x} \\
f_{\rm y} & f_{\rm x}-3
 \end{array}
 \right)
\left(
 \begin{array}{c}
 \delta y\\
 \delta x
  \end{array}
 \right),
\]
where
 \be
 R_{\rm y}\equiv\left.\partial R/\partial
 y\right|_{(y=y_c,x=x_c)},
 \ee
and the definitions of $R_{\rm x}$, $f_{\rm y}$, $f_{\rm x}$ and
$G_{\rm y}$ are similar. The functions $G$ and $R$ are defined by
 \[
 G=G(y)=\frac{4y}{2+y},
 \]
 \[
 R=R(x,y)=-\frac{f(x,y)}{(2+y)e^y},
 \]
which are used for the simplification of the notation. The two
eigenvalues of the coefficient matrix $M$ determine the stability
of the corresponding critical point. The critical point is an
attractor solution, which is stable, only if both the these two
eigenvalues are negative (stable node), or real parts of these two
eigenvalues are negative and the determinant of the matrix $M$ is
negative (stable spiral), which requires that the critical point
satisfies the following inequalities
 \be
 G_{\rm y}+R_{\rm y}+f_{\rm x}-3<0,\label{(1)}\ee
 \be
 \left[R_{\rm x}f_{\rm y}-(f_{\rm x}-3)(G_{\rm y}+R_{\rm y})\right]
 [(G_{\rm y}+R_{\rm y}-f_{\rm x}+3)^2-4R_{\rm x}f_{\rm y}]<0.\label{(2)}
 \ee
or it satisfies
 \be
 G_{\rm y}+R_{\rm y}+f_{\rm x}-3<0,\label{(3)}
 \ee
 \be
 (G_{\rm y}+R_{\rm y}-f_{\rm x}+3)^2-4R_{\rm x}f_{\rm y}=0.\label{(4)}
 \ee
These generate a constraint of the interaction term $Q$, which
will be shown in the following section.

Here we discuss some general features of the attractor solutions,
regardless the special form of the interaction term $Q$. From the
expression (\ref{[2]}), we find that $x_c=-\frac{4y_c}{3}e^{y_c}$.
Substitute this into the formula (\ref{Omega}), one obtains
 \be
 \Omega_y=\frac{(y_c+1)e^{y_c}}{(y_c+1)e^{y_c}+x_c}=\frac{3+3y_c}{3-y_c}.\label{Omegay}
 \ee
Since $0\leq\Omega_y\leq1$, this formula follows a constraint of
the critical point
 \be
 -1\leq y_c\leq0.\label{constrainty}
 \ee
From the formulae (\ref{13}) and (\ref{Omegay}), we obtain
 \be
 \Omega_yw_y=-1.\label{-1}
 \ee
This relation is kept for all attractor solutions, independent of
the special form of the interaction. Since the value of $\Omega_y$
is not larger than one in the attractor solution, we obtain that
 \be
 w_y\leq-1,
 \ee
the EOS of the YM dark energy must be not larger than $-1$,
phantom-like or $\Lambda$-like. Since in the early universe, the
value of the order parameter of the YM field $F$ is much larger
than that of $\kappa^2$, i.e. $y\gg1$, the YM field is a kind of
radiation component\cite{zhao2}. However, in the late attractor
solution, the dark energy is phantom-like or $\Lambda$-like. So
the phantom divide must be crossed in the former case, which is different from
the interacting quintessence, phantom or k-essence models.

In order to the investigate the finial fate of the universe, we
should investigate the total EOS in the universe, which is defined
by
 \be
 w_{tot}\equiv\frac{p_{tot}}{\rho_{tot}}=\frac{p_y+p_m}{\rho_y+\rho_m}=\Omega_yw_y,
 \ee
where $p_m=0$ is used. From the relation (\ref{-1}), we obtain
that, in the attractor solution,
 \be
 w_{tot}=-1.
 \ee
This result is also independent of the special form of the
interaction. So the universe is an exact de Sitter expansion, and
the cosmic big rip is naturally avoided, although the YM field
dark energy is phantom-like.


\section{Several interaction models}

In the previous section, we find the critical point of the
dynamical system of interacting YM dark energy models satisfies
not only the equations in (\ref{[1]}) and (\ref{[2]}), but also
the constraint of (\ref{constrainty}). It is obvious that the
expression of (\ref{[1]}) depends on the special form of
interacting term. If the critical point is an attractor, it also
satisfies the constraint in (\ref{(1)}) and (\ref{(2)}), or in
(\ref{(3)}) and (\ref{(4)}). These relations can give some
constraints of the interaction term. In this section, we consider
several cases with different interaction forms between the YM dark
energy and background matter, which are taken as the most familiar
interaction terms extensively considered in the
literature\cite{couple}.

\textbf{Case a}: $Q\propto H\rho_y$, which is equivalent to the
form $f(x,y)=\alpha(y+1)e^y$, where $\alpha$ is a dimensionless
constant. From the equations (\ref{[1]}) and (\ref{[2]}), we
obtain the critical point
 \be
 y_c=-\frac{\alpha}{4+\alpha},
 ~~x_c=-\frac{4y_c}{3}e^{y_c}.
 \ee
The constraint in (\ref{constrainty}) requires that
 \be
 \alpha\geq0,
 \ee
and the attractor conditions in (\ref{(1)})-(\ref{(4)}) require
that
 \be
 \alpha>-8.
 \ee
So we obtain the constraint of the interaction from, if the
attractor solution exists, the parameter $\alpha$ satisfies
 \be
 \alpha\geq0,
 \ee
and the EOS and the fractional energy density of the YM field in
the attractor solution are
 \be
 w_y=-\frac{1}{3}(\alpha+3),~~\Omega_y=\frac{3}{\alpha+3},
 \ee
respectively. It is obvious that $w_y\leq-1$.

\textbf{Case b}: $Q\propto H(\rho_y+\rho_m)$, which is equivalent
to the form $f(x,y)=\beta[(y+1)e^y+x]$, where $\beta$ is a
dimensionless constant. From the equations (\ref{[1]}) and
(\ref{[2]}), we obtain the critical point
 \be
 y_c=\frac{3\beta}{\beta-12},
 ~~x_c=-\frac{4y_c}{3}e^{y_c}.
 \ee
The constraint in (\ref{constrainty}) requires that
 \be
 0\leq\beta\leq3,
 \ee
and the attractor conditions in (\ref{(1)})-(\ref{(4)}) require
that
 \be
 \beta<\frac{120}{31}.
 \ee
So the parameter $\alpha$ satisfies
 \be
 0\leq\beta\leq3,
 \ee
if the critical state is an attractor solution. The EOS and the
fractional energy density of the YM field in the attractor are
 \be
 w_y=\frac{3}{\beta-3},~~\Omega_y=\frac{3-\beta}{3},
 \ee
respectively, which follows that $w_y\leq-1$, the
YM field dark energy is phantom-like or $\Lambda$-like.

\textbf{Case c}: $Q\propto H\rho_m$, which is equivalent to the
form $f(x,y)=\gamma x$, where $\gamma$ is a dimensionless
constant. From the equations (\ref{[1]}) and (\ref{[2]}), we easily Þnd that they have no 
solution except that the value of $\gamma$ is exactly zero, i.e. the case with no interaction.

\textbf{Case d}: Recently, a number of authors have discussed the
holographic dark energy, where the holographic principle has been
put forward to explain the dark energy. According to the
holographic principle, the number of degrees of freedom of a
physical system scales with the area of its boundary. In the
context, Cohen et al\cite{cohen} suggested that in quantum field
theory a short distant cutoff is related to a long distant cutoff
due to the limit set by formation of a black hole, which results
in an upper bound on zero-point energy density. In line with this
suggest, Hsu and Li\cite{Hsu,Li} argued that this energy density
could be views as the holographic dark energy satisfying
 \be \rho_{de}=3d^2M_P^2L^{-2}~,\label{de} \ee
where $d\geq0$ is a numerical constant, and $M_P\equiv
1/\sqrt{8\pi G}$ is the reduced Planck mass. $L$ is the size of
the current universe. Li\cite{Li} proposed that the IR cut-off $L$
should be taken as the size of the future event horizon

\be L=R_{eh}(a)=a\int_t^\infty{d\tilde{t}\over
a(\tilde{t})}=a\int_a^\infty{d\tilde{a}\over
H\tilde{a}^2}~.\label{eh} \ee

In this letter, we consider the holographic YM field dark energy.
From the relation (\ref{de}), one obtains that
 \be
 \dot{\rho}_y=\dot{\rho}_{de}=6M_p^2\Omega_yH^3\left(\frac{\sqrt{\Omega_y}}{d}-1\right),
 \ee
which follows that the interaction form is
 \be
 Q=-2\rho_yH\left(\frac{\sqrt{\Omega_y}}{d}-1\right)-3H\rho_y(1+w_y),
 \ee
where the expression (\ref{doty}) is used. This formula is
equivalent to the form
 \be
 f(x,y)=\left[-\frac{4y}{y+1}-2\left(\frac{\sqrt{\frac{3+3y}{3-y}}}{d}-1\right)\right](y+1)e^y.
 \ee
From the equations (\ref{[1]}) and (\ref{[2]}), we obtain the
critical point
 \be
 y_c=-\frac{3(d^2-1)}{3+d^2},
 ~~x_c=-\frac{4y_c}{3}e^{y_c}.
 \ee
The attractor conditions in (\ref{(1)})-(\ref{(4)}) require that
 \be
 d<0,
 \ee
which is conflicting with the previous assumption $d\geq0$. So we
get the conclusion: the holographic YM dark energy model has no
attractor solution.


\section{Conclusion and discussion}

In summary, the cosmological evolution of the Yang-Mill field dark
energy interacting with background matter is investigated in this
letter. We find the features of the interacting YM dark energy
models:

\emph{a.} The interaction term between the YM dark energy model
and the matter has a fairly tight constraint, if we require that
the attractor solution of the model exists.

\emph{b.} If the attractor solution exists, the EOS of the YM
field must evolve from $w_y>0$ to $w_y<-1$ or $w_y=-1$.

\emph{c.} The holographic YM dark energy model has no attractor
solution, which is different from other holographic
models\cite{holographic}.

\emph{d.} In the attractor solution, the total EOS is
$w_{tot}=-1$, which is independent of the interacting forms. So
the universe is in a de Sitter expansion, and the cosmic big rip
does not exist in the models.

In the interacting YM dark energy models, we should notice the
``fine-tuning" problem, which is reflected by the value of
$\kappa$, the energy scale of the Yang-Mills field dark energy
models. In the interacting models, the total energy density in the
universe is
 \be
 \rho_{tot}=\frac{\rho_m}{\Omega_m}=\frac{b\kappa^2}{2}\left[\left(1+y\right)e^y+x\right].
 \ee
In the attractor solution, we can obtain
 \be
 \rho_{tot}=\frac{b\kappa^2}{2}\left(1-\frac{1}{3}y_c\right)e^{y_c}.
 \ee
where the express (15) is used. The value of $\rho_{tot}$ should
be not larger than which of the present total energy density in
the universe\cite{seljak}, i.e.
 \be
 \rho_{tot}\leq8.099h^2\times10^{-11}eV^4,
 \ee
which leads to
 \be
 \kappa\leq4.18h\times10^{-5}eV^2\left(1-\frac{1}{3}y_c\right)^{-1/2}e^{-y_c/2}.
 \ee
For a fixed interacting models, where $y_c$ can be obtained, one
can exactly calculate the value of the $\kappa$, which keeps the current energy density of YM dark energy being current observed value.  From (47), we
find that this energy scale $\kappa$ is, as well as the case with free Yang-Mills field models, very low compared to the typical energy scales in particle physics.

~

~


{\bf Acknowledgement:}

This work is supported by Chinese NSF under grant Nos. 10703005 and 10775119.


\begin{thebibliography}{99}

\bibitem{sn}
A.G.Riess et al., Astron.J. {\bf 116} (1998) 1009;

S.Perlmutter et al., Astrophys.J. {\bf 517} (1999) 565;

J.L.Tonry et al., Astrophys.J. {\bf 594} (2003) 1;

R.A.Knop et al., Astrophys.J. {\bf 598} (2003) 102;

A.G.Riess et al., astro-ph/0611572;

W.M.Wood-Vasey et al., astro-ph/0701041.


\bibitem{map}
C.L.Bennett et al., Astrophys.J.Suppl. {\bf 148} (2003) 1;

D.N.Spergel et al., Astrophys.J.Suppl. {\bf 148} (2003) 175;

D.N.Spergel et al., Astrophys.J.Suppl. {\bf 170} (2007) 377.


\bibitem{sdss}
M.Tegmark et al., Astrophys.J. {\bf 606} (2004) 702,~Phys.Rev.D
{\bf 69} (2004) 103501;

A.C.Pope et al., Astrophys.J. {\bf 607} (2004) 655;

W.J.Percival et al., MNRAS {\bf 327} (2001) 1297.



\bibitem{coin}
P.Steinhardt, in `Critical Problems in Physics' ed. by V.L.Fitch
and D.R.Marlow (Princeton U.Press, 1997);

R.H.Dicke and P.J.E.Peebles, in `General Relativity: An Einstein
Centenary Survey', ed. by S.W.Hawking \& W.Israel (Cambridge U.
Press, 1979);

E.J.Copeland, M.Sami and S.Tsujikawa, Int.J.Mod.Phys.D {\bf 15}
(2006) 1753.

\bibitem{quint}
C.Wetterich, Nucl.Phys.B {\bf 302}, 668 (1988);~Astron.Astrophys.
{\bf 301} (1995) 321;

B.Ratra and P.J.E.Peebles, Phys.Rev.D {\bf 37} (1988) 3406;

R.R.Caldwell, R.Dave and P.J.Steinhardt, Phys.Rev.Lett. {\bf 80}
 (1998) 1582;




\bibitem{phantom}
R.R.Caldwell, Phys.Lett.B {\bf 545} (2002) 23;

S.M.Carroll, M.Hoffman and M.Trodden, Phys.Rev.D {\bf 68} (2003)
023509;

R.R.Caldwell, M.Kamionkowski and N.N.Weinberg, Phys.Rev.Lett. {\bf
91} (2003) 071301;

M.P.Dabrowski, T.Stachowiak and M.Szydlowski, Phys.Rev.D {\bf 68}
 (2003) 103519;

V.K.Onemli and R.P.Woodard, Phys.Rev.D {\bf 70} (2004) 107301.


\bibitem{k}
C.Armendariz-Picon, T.Damour and V.Mukhanov,  Phys.Lett.B {\bf
458} (1999) 209;

T.Chiba, T.Okabe and M.Yamaguchi, Phys.Rev.D {\bf 62} (2000)
023511;

C.Armendariz-Picon, V.Mukhanov and P.J.Steinhardt, Phys.Rev.D {\bf
63} (2001) 103510;

T.Chiba, Phys.Rev.D {\bf 66} (2002) 063514.

\bibitem{quintom}
B.Feng, X.L.Wang and X.M.Zhang, Phys.Lett.B {\bf 607} (2005) 35;

Z.K.Guo, Y.S.Piao, X.M.Zhang and Y.Z~Zhang, Phys.Lett.B {\bf 608}
 (2005) 177;

X.F.Zhang, H.Li, Y.S.Piao and X.M.Zhang, Mod.Phys.Lett.A {\bf 21},
 (2006) 231;

M.Z.Li, B.Feng and X.M.Zhang, JCAP {\bf 0512} (2005) 002;

H.Wei, R.G.Cai and D.F.Zeng, Class.Quant.Grav. {\bf 22} (2005)
3189;


M.Alimohammadi and H.M.Sadjadi, Phys.Rev.D {\bf 73} (2006)
083527;

H.Wei, N.Tang and S.N.Zhang, Phys.Rev.D {\bf 75} (2007) 043009;


W.Zhao and Y.Zhang, Phys.Rev.D {\bf 73} (2006) 123509;

W.Zhao, Phys.Lett.B {\bf 655} (2007) 97;

M.R.Setare and E.N.Saridakis, Phys.Lett.B {\bf 668} (2008) 177.





\bibitem{trans}
P.S.Corasaniti, M.Kunz, D.Parkinson, E.J.Copeland and B.A.Bassett,
Phys.Rev.D {\bf 70} (2004) 083006;

S.Hannestad and E.Mortsell, JCAP {\bf 0409} (2004) 001;

A.Upadhye, M.Ishak and P.J.Steinhardt, Phys.Rev. D {\bf 72} (2005)
063501 ;

J.Q.Xia, G.B.Zhao, H.Li, B.Feng and X.M.Zhang, Phys.Rev.D {\bf 74}
(2006) 083521.



\bibitem{vikman}
A.Vikman, Phys.Rev.D {\bf 71} (2005) 023515.



\bibitem{sc}
H.Wei and R.G.Cai, Phys.Rev.D {\bf 72} (2005) 123507;

H.Wei and R.G.Cai, Phys.Lett.B {\bf 634} (2006) 9;

H.Wei and R.G.Cai, Phys.Rev.D {\bf 73} (2006) 083002.


\bibitem{scaling}
I.Zlatev, L.Wang and P.J.Steinhardt Phys.Rev.Lett. {\bf 82} (1999) 896;

P.J.Steinhardt, L.Wang and I.Zlatev, Phys.Rev.D {\bf 59} (1999) 123504 (1999).



\bibitem{couple}
E.J.Copeland, A.R.Liddle and D.Wands, Phys.Rev.D {\bf 57} (1998)
4686;

L.Amendolam, Phys.Rev.D {\bf 60} (1999) 043501;

L.Amendolam, Phys.Rev.D {\bf 60} (1999) 043511;

Z.K.Guo, R.G.Cai and Y.Z.Zhang, JCAP {\bf 0505} (2005) 002;

Z.K.Guo and Y.Z.Zhang, Phys.Rev.D {\bf 71} (2005) 023501;

T.Damour and A.M.Polyakov, Nucl.Phys.B {\bf 423} (1994) 532;

W.Zimdahl and D.Pavon, Phys.Lett.B {\bf 521} (2001) 133;

L.P.Chimento, A.S.Jakubi, D,Pavon and W.Zimdahl, Phys.Rev.D {\bf
67} (2003) 083513;

R.G.Cai and A.Z.Wang, JCAP {\bf 0503} (2005) 002.



\bibitem{vector}
C.Armendariz-Picon, JCAP {\bf 0407} (2004) 007;

V.V.Kiselev, Class.Quant.Grav. {\bf 21} (2004) 3323;

C.G.Boehmer and T.Harko, Eur.Phys.J.C {\bf 50} (2007) 423;

C.Beck and M.C.Mackey, arXiv:astro-ph/0703364.




\bibitem{vector2}
K.Bamba and S.D.Odintsov, arXiv:0801.0954;

K.Bamba, S.Nojiri and S.D.Odintsov, arXiv:0803.3384;

T.S.Koivisto and D.F.Mota, arXiv:0805.4229.



\bibitem{z}
Y.Zhang, Phys.Lett.B {\bf 340} (1994) 18;

Y.Zhang, Chin.Phys.Lett.{\bf 14} (1997) 237.


\bibitem{Zhang}
Y.Zhang, Gen.Rel.Grav. {\bf 34} (2002) 2155.

\bibitem{zhao1}
W.Zhao and Y Zhang, Class.Quant.Grav. {\bf 23} (2006) 3405.

\bibitem{zhao2}
W.Zhao and Y.Zhang, Phys.Lett.B {\bf 640} (2006) 69;

W.Zhao and D.H.Xu , Int.J.Mod.Phys.D {\bf 16} (2007) 1735;

W.Zhao, Int.J.Mod.Phys.D {\bf 17} (2008) 1245.

\bibitem{xia}
Y.Zhang, T.Y.Xia and W.Zhao, Class.Quant.Grav. {\bf 24} (2007) 3309.

\bibitem{2-loop}
T.Y.Xia and Y.Zhang, Phys. Lett. B. {\bf 656} (2007) 19;

M.L.Tong, Y.Zhang and T.Y.Xia, arXiv:0809.2123.

\bibitem{3-loop}
S.Wang, Y.Zhang and T.Y.Xia, JCAP {\bf 10} (2008) 037.



\bibitem{seljak}
U.Seljak {\it et al.}, Phys.Rev.D {\bf 71} (2005) 103515;

U.Seljak, A.Slosar and  P.McDonald, JCAP {\bf 0610} (2006) 014.


\bibitem{pagels}
H.Pagels and E.Tomboulis, Nucl.Phys.B {\bf 143} (1978) 485.

\bibitem{adler}
S.Adler, Phys.Rev.D {\bf 23}  (1981) 2905;

S.Adler, Nucl.Phys.B {\bf 217} (1983) 3881.

\bibitem{Pol}
H.Politzer, Phys.Rev.Lett. {\bf 30} (1973) 1346;

D.J.Gross and F.Wilzcek, Phys.Rev.Lett. {\bf 30} (1973) 1343.

\bibitem{coleman}
S.Coleman and E.Weinberg, Phys.Rev.D {\bf 7} (1973) 1888.

\bibitem{parker}
L.Parker and A.Raval, Phys.Rev.D {\bf 60} (1999) 063512.


\bibitem{cohen}
A.Cohen, D.Kaplan and A.Nelson, Phys.Rev.Lett.{\bf 82} (1999)
4971.



\bibitem{Hsu}
S.D.H.Hsu, Phys.Lett.B {\bf 594} (2004) 13.



\bibitem{Li}
M.Li, Phys.Lett.B {\bf 603} (2004) 1.


\bibitem{holographic}
X.Zhang, Phys.Lett.B {\bf 648} (2007) 1;

X.Zhang, Phys.Rev.D {\bf 74} (2006) 103505;

H.Li, Z.K.Guo and Y.Z.Zhang, Int.J.Mod.Phys.D {\bf 15} (2006) 869;

M.R.Setare, Phys.Lett.B {\bf 642} (2006) 1.



\end{thebibliography}
\end{document}